\documentstyle[preprint,epsfig,aps,floats]{revtex}
\tightenlines

\begin{document}

\font\fortssbx=cmssbx10 scaled \magstep2
\hbox to \hsize{
\hbox{\fortssbx University of Wisconsin - Madison}
\hfill$\vcenter{\hbox{\bf MADPH-98-1095}
          \hbox{November 1998}}$ }

\vskip2cm

\centerline{\bf Muon Detection of TeV Gamma Rays
from Gamma Ray Bursts}
\vskip 0.1 cm
\centerline{J. Alvarez-Mu\~niz}
\centerline{\it Departamento de F\'\i sica de Part\'\i culas, Universidade de
Santiago,}
\centerline{\it E-15706 Santiago de Compostela, Spain}
\centerline{F. Halzen}
\centerline{\it Department of Physics, University of Wisconsin,
Madison, WI 53706}
\vskip 0.2 cm
\begin{abstract}
Because of the limited size of the satellite-borne instruments, it has not been possible to
observe the flux of gamma ray bursts (GRB) beyond GeV energy. We here show
that it is possible to detect the GRB radiation of TeV energy and above, by detecting the muon secondaries produced when the gamma rays shower in the Earth's
atmosphere. Observation is made possible by the recent commissioning of underground detectors (AMANDA, the Lake Baikal detector and MILAGRO) which combine a low muon threshold of a few hundred GeV or less, with a large effective area of $10^3$~${\rm m^2}$ or more. Observations will not only
provide new insights in the origin and characteristics of GRB, they also
provide quantitative information on the diffuse infrared background.
\end{abstract}

\newpage

\section{Muon burst astronomy.}

High energy gamma rays produce muons when interacting in the Earth's
atmosphere. These can be efficiently detected, and the direction of the parent  
gamma ray reconstructed, in relatively shallow underground ``neutrino"
detectors\cite{physrep} such as the now operating Antarctic Muon And Neutrino Detector Array (AMANDA)\cite{AMANDA} and Lake Baikal telescope\cite{Baikal}. These instruments are positioned at a modest depth of order 1~kilometer and are therefore sensitive to muons with energies of a few hundred GeV, well below the TeV thresholds of other deep underground detectors such as Superkamiokande and MACRO\cite{physrep}. They are therefore able to detect muons from primary gamma rays of TeV energy and above, with a very large effective telescope area of $10^3$~${\rm m^2}$, or more. They infer the photon direction by reconstructing the secondary muon track with degree-accuracy. Although muons produced by gamma rays from astronomical sources compete with a large background of atmospheric cosmic ray-induced muons, during the short duration of a GRB this background is manageable. Using the time stamp provided by satellite observation, the detector only integrates background over the very short time of the burst which is of order one second. Unlike air Cherenkov telescopes, muon detectors cover a large fraction of the sky with a large duty cycle, e.g.\ essentially 100\% efficiency for more than a quarter of the sky in the case of the AMANDA detector with a South Pole location.

In this paper we demonstrate how large area detectors operating with a few hundred GeV
muon threshold, or less, provide a unique window of opportunity for observing  
GRB. While the fluxes of TeV photons are reduced by one or more orders of magnitude compared to GeV photons observed with satellites, and while only one percent of the gamma rays will produce a detected secondary muon, observation of GRB is
possible because the detectors are four orders of magnitude larger than, for  
example, the EGRET instrument on the Compton Gamma ray Observatory\cite{egret}.

By the most conservative estimates, we predict order 1 muon per year correlated in time and direction with GRB for the operating detectors, and as high as 50 per year for the most conservative GRB flux estimate normalized to the observed diffuse GeV cosmic background\cite{vazquez,totani}. The unknown energetics of GRB above GeV may yield much higher rates; see, for instance, Ref.\cite{totani}. Interestingly, the muon count is similar for a single nearby burst at redshift $z=0.1$, and a flux of a few muons or more in a 1~second interval in coincidence with a gamma ray burst cannot be missed. For a cosmological distributions such an event occurs within $2 \sim 3$ years of observation, taking into account all observational constraints.

Failure to observe a signal within a few years would establish a cutoff on the GRB flux not
much above the GeV-sensitivity of the operating detectors, or point to an  
unexpectedly large infrared diffuse background absorbing the TeV gamma rays  
over cosmological distances. The two possibilities can be distinguished on the
basis of observation of the occasional nearby burst. There is also the possibility of cosmological evolution of the sources. The observations are obviously relevant to the proposal that GRB are the sources of the highest energy cosmic  
rays\cite{vietri}.

Also of interest here is the MILAGRO detector\cite{milagro} and the proposed  
HANUL experiment in Korea\cite{hanul}. Although of more modest size compared to  
``neutrino" telescopes, MILAGRO's muon threshold is only $1.5$~GeV because of  
its location at the surface. Sensitivity to gamma rays of lower energy results in an increased flux, and compensates for its smaller effective area. We  
should here point out that MILAGRO, as well as the other instruments  
discussed, have other capabilities to study GRB. For instance,
AMANDA has sensitivity to the MeV neutrinos produced in the initial
collapse\cite{jaczko}, and to neutrinos of TeV energy, and above\cite{AMANDA}.  
The MILAGRO detector, beyond counting muons, can efficiently reconstruct
gamma ray showers once their energy exceeds hundreds of GeV\cite{milagro}.

\section{Muons in Gamma Ray Showers}

When gamma rays interact with the atmosphere they initiate cascades of electrons and photons, but also some muons. The dominant source of muons is the decay of charged pions which are photoproduced by shower photons. The number of muons with energy above  
$E_{\mu}$ in a shower initiated by a photon of energy $E_{\gamma}$ was  
computed some time ago \cite{hikasa}. For $E_\mu$ in the range 0.1 to 1  
TeV the number of muons in a photon shower can be parametrized as
\begin{equation}
N_\mu(E_\gamma, >E_\mu)\simeq {2.14\times 10^{-5}\over \cos\theta}
{1\over (E_\mu/\cos\theta)} \left[{E_\gamma\over (E_\mu/\cos\theta)}\right],
\label{param}
\end{equation}
with energy in TeV units. The parametrization is valid for
$E_\gamma/E_\mu\geq 10$. This parametrization is adequate to calculate muon
rates in the AMANDA and Baikal detectors since both have muon thresholds of the
order of a few hundreds GeV.

For muons energies below 0.1 TeV, muon decay and muon energy losses in the
atmosphere must be taken into account. To accomplish this we have used a
linear shower Monte Carlo simulation which follows all shower particles down  
to muon threshold. It accounts for all decay modes and energy
losses\cite{gaisser}. We have calculated the number of muons at sea level with  
energy above 1.5 GeV, which is MILAGRO's threshold for muon detection. We  
have parametrized the results as a function of initial photon energy at  
different zenith angles for use in subsequent calculations.

The Monte Carlo calculates the production of muons by the decay of photoproduced  
pions. The photoproduction cross section is obtained by interpolation of  
$\gamma$-proton data, including the most recent high energy measurements\cite{databook}  
performed with accelerators. It is converted to a $\gamma$-air cross section using the  
empirical $A^{0.91}$ dependence on the atomic number. A constant diffractive  
cross section of 0.194~mb proceeding via $\rho$ production has been included  
in the Monte Carlo simulation. For high energy muons, the Monte Carlo results are adequately parametrized by Eq.~(1).

We here neglected the production of muons by  direct $\mu$-pair production and leptonic decay of charmed particles which only contribute to the flux of muons with energy above  
several TeV\cite{hikasa}.

\section{GRB Muon Rates from Energy Considerations}

It is thought that GRB are produced when a highly relativistic shock with
Lorentz factor of order 100 or higher dissipates its kinetic energy in  
collisions with the
inter-stellar medium or in internal collisions within the relativistic
ejecta. The observed $\gamma$-rays are most likely produced by synchrotron
radiation of electrons accelerated in the shock, possibly followed by inverse  
Compton scattering, when the relativistic shell becomes optically thin to pair  
production\cite{piran}.

The observed injection rate in the Universe is $\dot E_{\rm GRB}= 4\times
10^{44}~{\rm erg~Mpc^{-3}~yr^{-1}}$. Interestingly, this is equal to the injection in cosmic  
rays beyond the ankle in the spectrum near $10^6$ TeV\cite{vietri}. The BATSE
satellite observes, on average, 1 GRB per day with an efficiency of about
$25\%$. Therefore the average energy emitted per burst, $E_{\rm GRB}$, is :
\begin{equation}
E_{\rm GRB}\simeq 3\times 10^{52}~{\rm ergs}
 \left({D\over 3000~{\rm Mpc}}\right)^3
\left({\dot E\over 4\times 10^{44} {\rm erg~Mpc^{-3}~yr^{-1}}}\right)
\left({4~{\rm day}^{-1}\over R}\right).
\label{energy}
\end{equation}
$R$ is the burst rate. Throughout our analysis we will assume a cosmological distribution of  GRB, with a distance of $D=3000$ Mpc to the average burst. Assuming no beaming, this energy corresponds to an average fluency per GRB, $F_{\rm GRB}$, close to the one observed:
\begin{equation}
F_{\rm GRB}=3\times 10^{-8}~{\rm J~m^{-2}}
\left({3000~{\rm Mpc}\over D}\right)^2
\left({E_{\rm GRB}\over 3\times 10^{52}~{\rm ergs}}\right).
\end{equation}
Although in line with observations, this may be an underestimate because present detectors provide no information on the energetics of GRB above GeV-energy. Some authors \cite{vazquez,totani} have, for instance, raised the interesting possibility that GRB are the origin of the diffuse extragalactic gamma ray background. This association requires a photon energy of order $E_{\rm GRB}\sim 10^{54}$--$10^{56}$ ergs per burst, depending on the GRB occurrence rate assumed.  
Most of this energy is concentrated in the high energy tail of the photon
spectrum, above $E_\gamma>1$ TeV, and is therefore not included in the energy balance of equation~\ref{energy} where only observed photons of GeV-energy and below have been considered. There are models that can accommodate the
large total energy required by this scenario, for instance those where a complete neutron star is converted into photons\cite{turok}.

A typical GRB spectrum exhibits a high energy tail which, above a few  
hundreds keV, can be parametrized by a power law:
\begin{eqnarray}
{dN_\gamma\over dE_\gamma}=
{F_\gamma\over E^{(\gamma +1)}}~10^{-12}~{\rm cm^{-2}~s^{-1}\;,}
\end{eqnarray}
where energies are in TeV. Typical values for the observed spectral index
$\gamma$ range from 0.8 to 1 \cite{piran}. The photon spectrum of the average  
burst is normalized by energy conservation:
\begin{equation}
\int_{E_{\gamma {\rm min}}}^{E_{\gamma {\rm max}}} dE_\gamma~
E_\gamma~{dN_\gamma\over dE_\gamma}=
{F_{GRB}\over \Delta t}\;,
\end{equation}
where $\Delta t$ is the average duration of the burst of order 1 second.
$E_{\gamma {\rm min}}$ ($E_{\gamma {\rm max}}$) is the minimum (maximum)
energy of the photons emitted by the burst. For $E_{\gamma {\rm min}}$ we will  
use $1~{\rm MeV}$ throughout. Observations indicate indeed that a negligible  
fraction of the energy is emitted in X-rays.

The TeV flux obtained by  
extrapolation, is too low to be
detected by the present satellite experiments because of their small telescope area. The high  energy behavior of the GRB spectrum is therefore a matter of intense speculation. Bursts  
where isolated photons reach tens of GeV energy have been detected, suggesting  
the extension of the spectrum to the TeV range\cite{piran}. As previously
discussed, the extension of  the photon spectrum to TeV energy with a
rather flat spectral index ($\gamma\sim$ 0.5) is an essential feature of
models accommodating the diffuse GeV background. Also, the HEGRA group\cite{HEGRA} has observed an excess of
gamma rays with $E_\gamma>$16 TeV in
temporal and directional coincidence with GRB 920925c.

The value of $F_\gamma$ obtained through above normalization procedure is given by:
\begin{eqnarray}
F_\gamma&=&3\times 10^4~
\left({3000~{\rm Mpc}\over D}\right)^2
\left({E_{\rm GRB}\over 3\times 10^{52}~{\rm ergs}}\right)\nonumber\\
&&\quad \times \cases{(1-\gamma)~[E_{\gamma {\rm max}}^{(1-\gamma)}
-E_{\gamma {\rm min}}^{(1-\gamma)}]^{-1}~~{\rm for}~~\gamma\ne 1~\cr
\left[ln\left({E_{\gamma{\rm max}}\over E_{\gamma{\rm min}}}\right)\right]^{-1}
~~{\rm for}~~\gamma=1}
\label{fgamma}
\end{eqnarray}
Note that, to a first approximation, the normalization
factor $F_\gamma$ only depends on $E_{\gamma {\rm min}}$ ($E_{\gamma {\rm
max}}$) for $\gamma >1$ ($\gamma <1$). The reason for this is clear, when
$\gamma >1$ the spectrum is very steep and most of the GRB energy is
concentrated in lower energy photons. The situation is reversed in the case of  
$\gamma <1$. We will explore the dependence of the muon rate on $\gamma$
further~on.

Absorption of gamma rays in the infrared, optical and microwave backgrounds
is a determining factor in the gamma ray flux observed at Earth. The mean  
free path of a TeV photon in the diffuse background is thought to be less than a  
few hundred Mpc, although this estimate is very uncertain because of our poor knowledge of the diffuse infrared background. Nevertheless,  
for a cosmological population of astrophysical objects such as GRB, absorption is expected to reduce the detected flux of TeV photons. The detection could therefore be
dominated by the closest bursts.

We first compute $N_\mu(E_\gamma,>E_\mu)$, the number of muons with energy in  
excess of $E_\mu$, produced in a  photon shower of energy $E_\gamma$. The muon rate is  
subsequently derived by convolution with the gamma ray spectrum,
\begin{equation}
N_\mu(>E_\mu)=\int_{E_{\gamma {\rm th}}}^{E_{\gamma {\rm max}}} dE_\gamma~
N_\mu(E_\gamma,>E_\mu)~{dN_\gamma \over dE_\gamma}\;.
\end{equation}
Here $E_{\gamma {\rm th}}$, the minimum photon energy needed to
produce muons of energy $E_\mu$, is given by
$E_{\gamma {\rm th}}\sim 10\times E_\mu/cos\theta$, where $\theta$ is
the zenith angle at which the source is observed.

Our results imply that the operating AMANDA, Baikal and MILAGRO detectors can do GRB science in an energy range not covered by astronomical telescopes. While detection may be marginal for the most conservative estimates of the high energy GRB flux, this will not be the case for future detectors on the drawing board such as AMANDA II, IceCube\cite{AMANDA} and Antares\cite{antares}.

\subsection{AMANDA and Lake Baikal as Gamma Ray Telescopes}

The vertical threshold for muon detection in the AMANDA detector at a 1.5~km  
depth is
$\sim 350$~GeV. For the shallower depth of lake Baikal this threshold is $\sim
150$ GeV. For threshold energies in this range we may make use of parametrization  
(\ref{param}) which is adequate for $E_\mu$ between 0.1 and 1 TeV\cite{yodh}.  
As previously mentioned, absorption of gamma rays in cosmological backgrounds  
has to
be taken into account, as well as the source distribution of GRB in redshift $z$:
\begin{eqnarray}
N_\mu (>E_\mu) ({\rm yr^{-1}})=A({\rm m^{2}})~\Delta~t~{1\over 2}~
t~\int_0^{\theta_{max}}~\int_0^{z_{max}}
\int_{E_{\gamma{\rm min}}}^{E_{\gamma{\rm max}}}~dz~dE_\gamma
{F_\gamma \over E_\gamma^{\gamma+1}}~e^{-\tau(E_\gamma,z)} \nonumber\\
{}\times{2.14\times 10^{-17} \over E_\mu}
\left( E_\gamma \cos\theta \over E_\mu \right) \sin\theta 
{dR_{GRB}\over dz}~(z)\;,
\end{eqnarray}
where $E_{\mu}$ is the vertical threshold energy of the detector, in TeV units.   
$\tau(E_\gamma,z)$ is the
optical depth of a photon with energy $E_\gamma$ originating at a distance  
$z$. For illustration, we will use the diffuse photon background of reference
\cite{stecker} throughout. $dR_{\rm GRB}/dz$ is the cosmological distribution of GRB,  
i.e.\ the number of GRB per unit time and distance $z$. Following  
\cite{mannheim,piran} we will assume cosmologically distributed standard candles with no  
source evolution. The GRB distribution has been normalized to 365 events per  
year observed by BATSE below a redshift of $z_{\rm max}=2.1$\cite{mannheim}. $A$ is the effective area of the detector. The dependence of $F_\gamma$ on the  
characteristics of the burst is given by Eq.(\ref{fgamma}), which can  
be used for scaling purposes.

AMANDA and Baikal observations require a time-stamp for background rejection,  
they can therefore only see bursts previously observed by satellite
experiments. The BATSE rate of 1 burst/day corresponds to an isotropic
distribution on the sky of 365 bursts/year. For the most conservative estimate with $E_{\gamma {\rm max}}=10$ TeV and a ``standard" burst with total energy $E_{\rm GRB}=3\times10^{52}$ ergs and a spectral index $\gamma = 1$ , we predict 1 muon per year correlated in time and direction with GRB for the AMANDA telescope, and 89 per year for the most conservative GRB flux estimate normalized to the observed diffuse GeV cosmic background\cite{vazquez,totani}. The unknown energetics of GRB above GeV may yield much higher rates; see, for instance ref\cite{totani}. The results are subject to the uncertainty associated with the absorption on infrared light which we computed following Stecker\cite{stecker}. The rates are essentially the same for the Lake Baikal experiment.

Interestingly, the muon counts are similar for a single nearby burst at redshift $z=0.1$, and a flux of a few muons or more in a 1~second interval in coincidence with a gamma ray burst provides a striking experimental signature which is hard to miss. In Figs.~1 and 2 we plot the number of muons per burst versus spectral index $\gamma$ for a burst at $z=0.1$. We assume again that the photon spectrum only extends to $E_{\gamma {\rm max}}=10$ TeV with a total energy $E_{\rm GRB}=3\times10^{52}$ ergs. We assumed an effective area of $A=10^4~{\rm m^2}$ for AMANDA  
and $10^3~{\rm m^2}$ for the Lake Baikal detector\cite{AMANDA,Baikal}. The results are shown with and without absorption on infrared light. Also the dependence of the
rate on zenith angle $\theta$ is illustrated. The rapid decrease of the number  
of events when $\theta$ increases is a consequence of the increase of the
energy threshold because muons have to penetrate an increasing amount of
matter to reach the detector. The smaller area of Baikal is compensated by its  
lower muon threshold, especially for larger zenith angles.

In Fig.~3 we show the rate per burst in AMANDA detector
as a function of redshift for a model with spectral index $\gamma=1$ and  
total energy $E_{\rm GRB}=3\times 10^{52}$ ergs. Also shown is the flux for  
energetics\cite{vazquez} which accommodates the diffuse GeV background.  
Absorption has been included, as in Figs.~1 and 2. Shown as a dashed line is  
the number of days towards a burst at the corresponding redshift assuming the  
GRB distribution of reference \cite{mannheim}.

An important question is whether the signal is observable given the large
background of muons of cosmic ray origin penetrating the detectors. In this
respect the great advantage of GRB detection is that the detector only
integrates background for the short duration of the burst, typically 1 second.  
Furthermore, the background can be limited to a circle in the sky of angle
$\delta \theta$ around the direction of the burst. Here $\delta \theta$ is the  
angular resolution of the detector which is typically a few degrees. It can be  
sharpened up by quality cuts, but this inevitably results in a loss of  
effective area. The muon background intensity from cosmic ray showers  
$I_\mu(\theta)$ has been measured\cite{AMANDA,Baikal}. The muon background at zenith $\theta$ is given by:
\begin{equation}
{\rm Noise}=I_\mu(\theta)\times A\times \delta\theta^2\;.
\end{equation}
The signal-to-square root of noise then scales as
\begin{equation}
{{\rm S}\over \sqrt {\rm N}}={\sqrt A\over \delta\theta}\;,
\end{equation}
which simply expresses that the sensitivity is improved for larger area and
for better angular resolution. Note that our search is not sensitive to
photons produced over time-scales much larger than seconds, for instance to
photons possibly produced by propagation of the external shock in the interstellar medium.  

In Table I we present the number of muons per burst detected by AMANDA. The
first column is the redshift of the burst, the second (third) column is the signal without (with) absorption of gamma rays in
intergalactic backgrounds. In each columns the results is given for zenith angle $\theta= 60$ and $\theta= 0$. The fourth column gives the signal to square root of  
noise corresponding to column 3. The 5th column shows the time in days between  
bursts with $z$ lower than the value shown in the first column. We here
assumed a ``standard" burst with spectral index $\gamma=0.8$ and $E_{\gamma
{\rm max}}=10$ TeV.

\begin{center}
\begin{tabular}{|c|c|c|c|c|} \hline
 $z$ & Muons & Muons &Signal to & Time \\
  & per burst & per burst & $\sqrt{\rm Noise}$ & (days) \\ \hline\hline
 0.05 & 5 -- 25 &  2 -- 12  & $60\over\delta\theta$ -- $133\over\delta\theta$ & 8400 \\
 0.1 & 1 -- 6 & 0.06 -- 0.6 & $1.8\over\delta\theta$ -- $6.6\over\delta\theta$ &   1200 \\
 0.5 & 0.05 -- 0.25 & $5\times 10^{-10}$ -- $3\times 10^{-6}$ &
${1.5\times 10^{-8}\over\delta\theta}$
-- ${3.3\times 10^{-5}\over\delta\theta}$ & 24 \\
 1.0 & 0.01 -- 0.05 & 0 -- $3\times 10^{-11}$  &
0 -- ${3.3\times 10^{-10} \over\delta\theta}$ & 8  \\ \hline
\end{tabular}
\label{grbamanda}
\end{center}

\noindent
{\bf Table I:} {\narrower Event rates in AMANDA detector. For models accommodating the diffuse GeV background the rates are larger by 2 orders of magnitude. The
first column is the redshift of the burst, the second (third) column is the signal without (with) absorption of gamma rays in intergalactic backgrounds. In each columns the results is given for zenith angle $\theta=$ 60 and $\theta=$ 0. The fourth column gives the signal to square root of noise corresponding to column 3. The 5th column shows the time in days between  
bursts with $z$ lower than the value shown in the first column.\par}

\bigskip\noindent
The role of absorption for gamma ray detection is evident. Nearby bursts with  
$z<0.1$ provide the best opportunity for detection. Their frequency is of the  
order of 1 burst every $2 \sim 3$ years.

\subsection{The MILAGRO Telescope}

The muon energy threshold for MILAGRO detector is only 1.5 GeV. It has an
effective area $A=1.5\times 10^3~{\rm m^2}$ and an intrinsic angular
resolution of about $3^{\circ}$. This requires the use of a 4.7$^{\circ}$
bin-size to collect $\sim$~70\% of the gamma ray events in which the muon
background from cosmic rays is $\sim$ 900 Hz \cite{milagro}.
In Table~II we present the number of muons per
burst observed by MILAGRO for two maximum energies
and two spectral slopes of the photon spectrum. The results are shown
corresponding to the ``standard" burst energy, and to the energy  
required in the models where GRBs are responsible for the diffuse  
extragalactic gamma ray background.
The absorption of the photon flux has not been taken into account. MILAGRO  
with a threshold for muon detection as low as 1.5 GeV is sensitive
to photons with energy $\sim$15 GeV which are not absorbed in the
intergalactic backgrounds.
Therefore detection of bursts at large
redshifts is possible.
For a spectral index $\gamma=1$, equal amounts of energy are stored in
every logarithmic interval. The number of low energy photons will be larger than high energy ones by the ratio of energies. On the other hand, the number of muons in a photon shower scales roughly with its energy. This will compensate the contribution to the number of muons
from low and high energy photons if it were not for absorption which predominantly reduces the high energy part of the photon spectrum. Counting of low energy muons may therefore be  
more efficient than reconstruction of the photon shower, the conventional MILAGRO method, because it raises the threshold to hundreds of GeV.  This is especially true for
the detection of the average ($z\sim 1$) GRB.

\begin{center}
\begin{tabular}{|c|c|c|c|c|} \hline
 $E_{\rm GRB}$ & Spectral Slope & $E_{\gamma {\rm max}}$ &
Signal & Signal to \\
(ergs) & $\gamma$ & (TeV)  & per burst & $\sqrt{\rm Noise}$ \\ \hline\hline
$3\times 10^{52}$ & 1.0 & 10. & 24 & 0.82 \\
$3\times 10^{52}$ & 0.8 & 10. & 57 & 1.9 \\
$3\times 10^{52}$ & 0.5 & 10. & 95 & 3.2 \\ \hline
$3\times 10^{52}$ & 1.0 & 0.1 & 2.3 & 0.08 \\
$3\times 10^{52}$ & 0.8 & 0.1 & 5.0 & 0.16 \\
$3\times 10^{52}$ & 0.5 & 0.1 & 9.2 & 0.3 \\ \hline
$10^{54}$ & 1.0 & 10. & 785 & 26.4 \\
$10^{54}$ & 0.8 & 10. & 1920 & 64.0 \\
$10^{54}$ & 0.5 & 10. & 3230 & 107.4 \\ \hline
$10^{54}$ & 1.0 & 0.1 & 75 & 2.5 \\
$10^{54}$ & 0.8 & 0.1 & 165 & 5.6 \\
$10^{54}$ & 0.5 & 0.1 & 305 & 10.0 \\ \hline
\end{tabular}
\label{grbmilagro}
\end{center}

\noindent
{\bf Table II:} {\narrower Sensitivity of MILAGRO to a GRB at
zenith angle $\theta$=0, for several assumptions on total
energy, spectral index and
maximum energy.
Absorption of gamma rays is not included.\par}

\vskip 1 cm 
\noindent{\bf Acknowledgements:}

J.A. thanks the Phenomenology Institute at the University of Wisconsin,
Madison for its hospitality. We thank Steve Barwick, Enrique Zas and Ricardo Vazquez for discussions. The stay of J.A. at the Univ. of Wisconsin, Madison,
was supported by the Xunta de Galicia.
This research was supported in part by the U.S.~Department of Energy under Grant No.~DE-FG02-95ER40896 and in part by the University of Wisconsin Research Committee with funds granted by the Wisconsin Alumni Research Foundation.

\newpage

\begin{figure}
\centering
\leavevmode
\epsfig{file=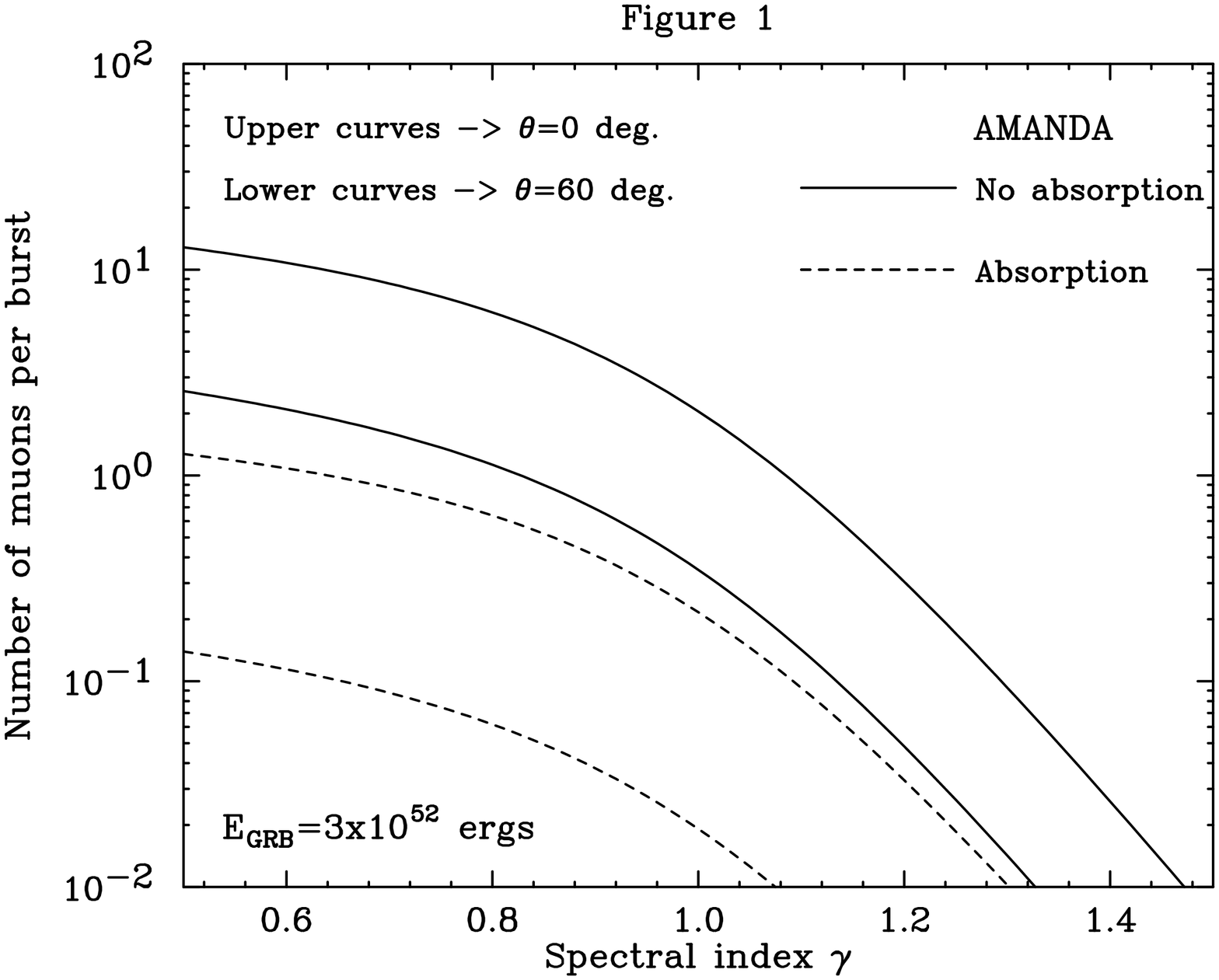,angle=90,width=6in}
\bigskip
\caption{Number of muons in AMANDA per ``standard" burst (see text) at
$z=0.1$ as a function of the spectral index of GRB photon spectrum.
Curves are shown for different zenith angles of observation and with and
without absorption of photons in the infrared background.}
\end{figure}

\begin{figure}
\centering
\leavevmode
\epsfig{file=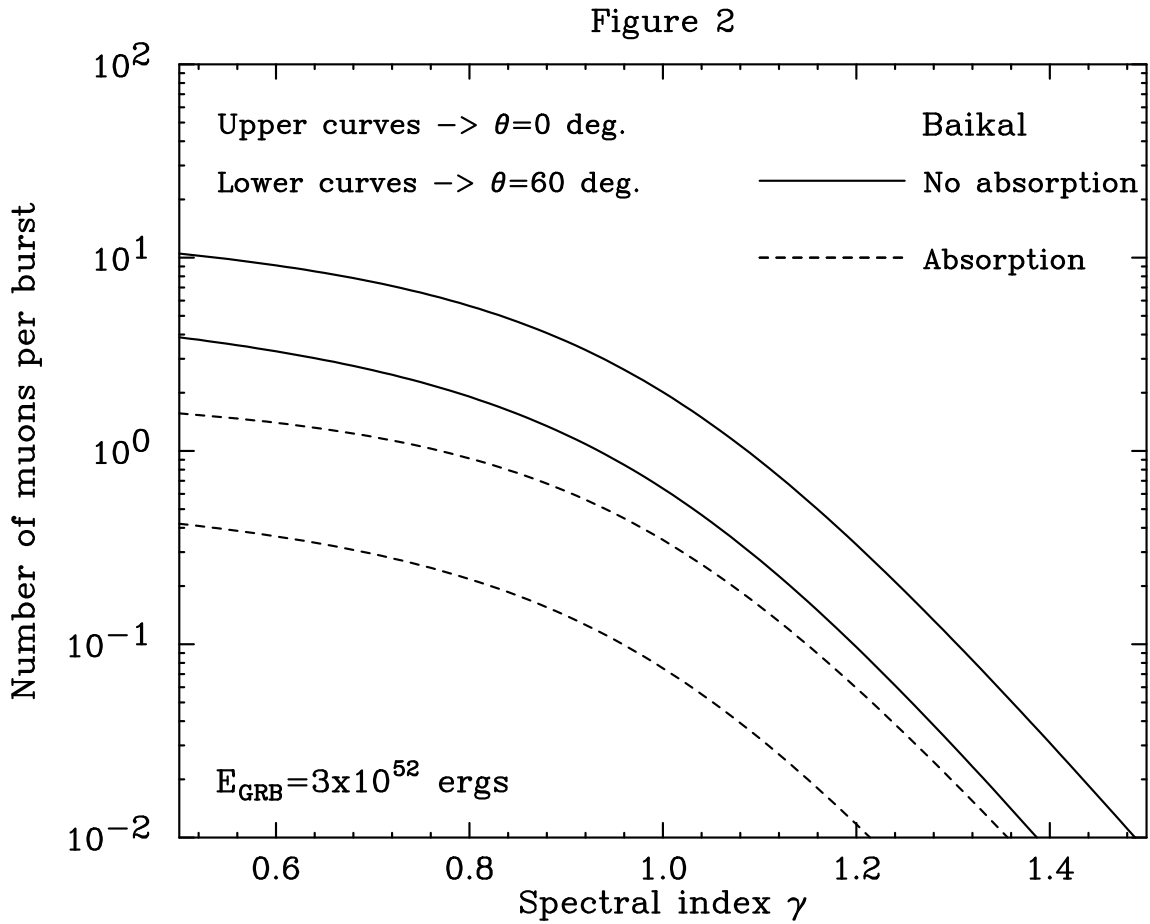,angle=90,width=6in}
\bigskip
\caption{Same as Fig.~1 for Lake Baikal detector.}
\end{figure}

\begin{figure}
\centering\leavevmode
\epsfig{file=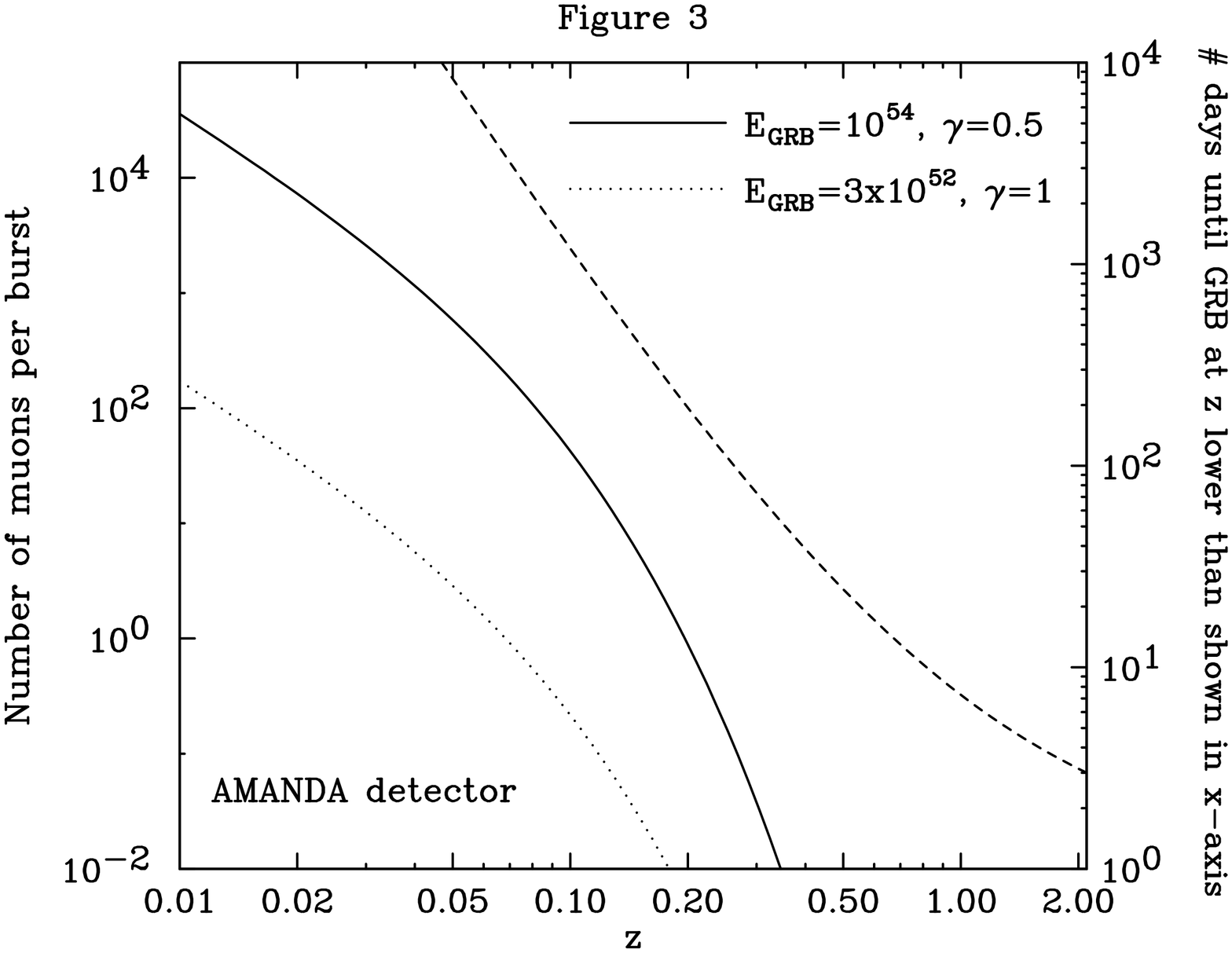,angle=90,width=6in}
\bigskip\leavevmode
\caption{Number of muons per burst in AMANDA as a function of redshift.
Results include absorption by the infrared background and are shown for a
``standard" burst and a burst normalized to the observed GeV diffuse
gamma ray background.
The dashed line shows the time in days until a burst occurs in the field of view of AMANDA, at a $z$ value lower than the one indicated on the horizontal axis.}
\end{figure}


\begin{thebibliography}{999}
%
\bibitem{physrep} T.K. Gaisser, F. Halzen, T. Stanev, {\it Phys. Rep.} {\bf
258}, 173 (1995) and references therein.
%
\bibitem{AMANDA} AMANDA collaboration, hep-ex/9809025,
talk presented at the 18th International Conference on Neutrino Physics
and Astrophysics (Neutrino 98), Takayama, Japan, June 1998.
%
\bibitem{Baikal} BAIKAL Collaboration, talk presented at the 18th International Conference on Neutrino Physics and Astrophysics (Neutrino 98), Takayama, Japan, June 1998; I.~Sokalski and C.~Spiering, The BAIKAL Neutrino Telescope NT-200, BAIKAL 92-03 (1992); C. Spiering, astro-ph/9801044, talk given at {\it Int. School of Nuclear Physics}, Erice, Sept.~1997.
%
\bibitem{egret}
D.\,J.\,Thompson {\it et al.}, {\it Ap. J. S} {\bf 101}, 259 (1995).
%
\bibitem{vazquez} R.A. V\'azquez, astro-ph/9810231.
%
\bibitem{totani} T. Totani, astro-ph/9810207.
%
\bibitem{vietri} E. Waxman, Phys. Rev. Lett. {\bf 75}, 386 (1995); M. Vietri, Phys. Rev. Lett. {\bf 80}, 3690 (1998) and references therein.
%
\bibitem{milagro} Gaurang B. Yodh, MILAGRO internal memo, 1995 (unpublished).
See also the MILAGRO home page, {\it http://umauhe.umd.edu/milagro.html}
%
\bibitem{hanul} W. Lee, {\it Nucl. Phys. B (Proc. Suppl.)} {\bf 66}, 252 (1998).
%
\bibitem{jaczko} F.\,Halzen and G.\,Jaczko, {\it Phys. Rev. D} 
{\bf 54}, 2774 (1996).
%
\bibitem{hikasa} F. Halzen, K. Hikasa and T. Stanev, {\it Phys. Rev. D}  {\bf 34},
2061 (1986) and references therein.
%
\bibitem{gaisser} T.K. Gaisser, {\it Cosmic Rays and Particle Physics},
Cambridge University Press, Cambridge, England, 1990.
%
\bibitem{databook} C. Caso {\it et al}, {\it Eur. Phys. J.} {\bf 3}, 1 (1998).
%
\bibitem{piran} For a recent review on GRB see T. Piran, astro-ph/9810256
(to appear in {\it Phys. Rep.})
and references therein.
%
\bibitem{turok} U. Pen, A. Loeb and N. Turok, astro-ph/9712178.
%
\bibitem{HEGRA} L. Padilla {\it et al.} (HEGRA collaboration),
astro-ph/9807342, accepted for publication in {\it Astron. and Astroph.}
%
\bibitem{antares} C. Arpesella, {\it Nucl. Instrum. Meth. A} {\bf 409},
454 (1998).
%
\bibitem{yodh} F. Halzen, T. Stanev and G.B. Yodh, {\it Phys. Rev. D} {\bf 55},
4475 (1997).
%
\bibitem{stecker} F.W. Stecker and O.C. de Jager, {\it Astron. and Astroph.}
{\bf 334}, L85 (1998).
%
\bibitem{mannheim} K. Mannheim, D. Hartmann, B. Funk, {\it Ap. J.}
{\bf 467}, 532 (1996).
%
%
%
%
%
\end{thebibliography}
\end{document}